\documentclass[aip,reprint]{revtex4-1}
\usepackage{amsmath}
\usepackage{braket}
\usepackage{graphicx}
\usepackage{comment} 

\draft 

\begin{document}


\title{MixPI: Mixed-Time Slicing Path Integral Software for Quantized Molecular Dynamics Simulations} 



\author{Britta A. Johnson}
\email[]{britta.johnson@pnnl.gov}
\affiliation{
Physical Science Division, Pacific Northwest National Laboratory, Richland, Washington 99352, USA
}%
\author{Siyu Bu}%
\affiliation{ 
Department of Chemistry and Chemical Biology, Cornell University, Ithaca, New York 14850, USA 
}%
\author{Christopher J. Mundy}
\affiliation{
Physical Science Division, Pacific Northwest National Laboratory, Richland, Washington 99352, USA
}%
\author{Nandini Ananth}%
\email[]{ananth@cornell.edu}
\affiliation{ 
Department of Chemistry and Chemical Biology, Cornell University, Ithaca, New York 14850, USA 
}%


\date{\today}

\begin{abstract}
 Path Integral Molecular Dynamics (PIMD) is a well established simulation technique to compute exact equilibrium properties for a quantum system using classical trajectories in an extended phase space. Standard PIMD simulations are numerically converged by systematically increasing the number of classical ‘beads’ or replicas used to represent each particle in the quantum system. Currently available scientific software for PIMD simulations leverage the massively parallel (with respect to number of beads) nature of the classical PIMD Hamiltonian. For particularly high-dimensional systems, contraction schemes designed to reduce the overall number of beads per particle required to achieve numerical convergence are also frequently employed. However, these implementations all rely on using the same number of beads to represent all atoms/particles, and become inefficient in systems with a large number of atoms where only a handful contribute significant quantum effects. Mixed time slicing (mixTS) offers an alternate path to efficient PIMD simulations by providing a framework where numerical convergence can be achieved with different numbers of beads for different types of atoms. Unfortunately, mixTS is not available in existing PIMD software. In this paper, we introduce MixPI for atomistic mixTS-PIMD simulations within the open-source software package CP2K. We demonstrate the use of MixPI in two different benchmark systems: we explore the use of mixTS in computing radial distributions functions for water, and in a more significant demonstration, for a solvated Co$^{2+}$ ion represented as a classical Co$^{3+}$ ion in water with an explicit, quantized 1024-bead electron localized on the metal ion.
\end{abstract}


\maketitle 


\section{Introduction}\label{sec:into}

    Nuclear quantum effects (NQEs) play a key role across many chemical phenomena. These effects become pronounced near electronic state crossings, when studying light atoms and particles, and at low temperatures. 
    NQEs, like tunneling and nuclear zero-point energy, play a critical role in charge transfer reactions,\cite{Waluk2024,DeGregorio2020,Chu2022,Kenion2016,Pierre2017} the structure and dynamics of liquids and interfaces, \cite{Remsing2023,Chandler,Kuharski1985,Lamaire2019,Lan2022,Meier2018,Kurapothula2022,Cassone2020} and the behavior of species under confinement.\cite{Lamaire2019,Wahiduzzaman2014,Cendagorta2016,Kumar2006,Gao2019,Shrestha2019,Ganeshan2022,Bocus2023} A variety of methods have been developed to include NQEs in simulations including  exact quantum dynamics,~\cite{MEYER1990,BECK20001,Lopreore1999,Makri1995,Makri1998} Gaussian wavepackets,~\cite{Richings2015,Christopoulou2021} multi-component quantum theories,~\cite{Zhao2020,Zhao2020_2,Ishimoto2009,Hasecke2023} and semiclassical methods. \cite{Mayera2008,Wang1998, Miller2001, Malpathak2022} However, these methods are not generally
    applicable to high-dimensional condensed phase systems. 

    The path integral formulation of quantum mechanics offers a particularly attractive alternative for atomistic simulations with methods like path integral molecular dynamics (PIMD) capturing exact equilibrium properties for quantum systems employing only classical trajectories in an extended phase space.~\cite{Ceperley1995,Mittal2020,Ceriotti2010} In addition, approximate real-time methods based on PIMD like centroid molecular dynamics (CMD)~\cite{Cao1994}  and ring polymer molecular dynamics (RPMD) \cite{Craig2005, Habershon2011} have found broad application in the calculation of dynamic observables. 
    These methods exploit the classical isomorphism between the static properties of a quantum system and a classical `ring polymer' that consists of $N$ copies of the system (called beads) connected by harmonic springs.\cite{Feynman1965} The number of beads is typically determined through numerical convergence of a given simulation and can vary with temperature, atomic masses, and the observable 
    being computed. 
    
    Several open-source PIMD/RPMD software packages have been previously developed \cite{Kapil2019,ipi-3.0,cp2k,rpmd-lammps,openmm} and are commonly used to include NQEs in atomistic systems. 
    Even though PI methods scale similar to classical methods with respect to system size, the increased number of particles in a PI simulation from discretization can make the calculation $N$ times more expensive than a corresponding classical simulation, where $N$ is the total number of ring polymer beads.\cite{Markland2008} 
    While some of this cost can be mitigated through massively parallel implementations that leverage the structure of the PIMD Hamiltonian, systems with a large number of atoms can pose a significant challenge.

    Several numerical approximations have been developed to reduce the computational expense of PIMD simulations. For instance, ring polymer contraction methods separate the forces into short range (quickly varying) and long range (slowly varying) forces, and employ a larger number of beads for the short range forces while the long range forces are computed for a smaller number of beads. \cite{Markland2008,Tuckerman1992,Kapil2016} In addition, using neural net potentials and machine learning methods that lower the cost of the force evaluation step can significantly reduce PIMD simulation times. \cite{Kimizuka2022,Loose2022,Yao2021,Hellstrom2018, Brezina2023, Bocus2023, Mazo2021} PI coarse-graining offers yet another way to reduce overall system dimensionality and hence PIMD simulation costs while still accurately capturing key interactions necessary to model the underlying physics.\cite{Ryu2019,Ryu2022,Lawrence2023,Peng2014,Musil2022}

    Mixed-time slicing (mixTS) is a method that addresses the scaling problem of PIMD by numerically converging the simulation while allowing different numbers of beads for different atoms in the system.~\cite{Steele2011} Note that we use the acronym mixTS to distinguish from the multi-time slicing (MTS) approach used to separately treat short-range and long-range interactions in the PIMD framework.~\cite{Tuckerman1992}
    System-specific mixTS implementations have been previously
    reported in the literature, however, to the best of our knowledge, a general PIMD software that allows for 
    the treatment of atomistic systems in the quantum-classical 
    mixTS regime is not currently available. We expect mixTS will be particularly important in the simulation of systems with a very large number of atoms where only a handful contribute significantly to observable nuclear quantum effects. For instance, previous work has demonstrated the efficacy of a quantum-classical PIMD simulation, where a single quantum particle is discretized using $N$-beads while all the remaining particles/atoms are treated classically ( 1-bead limit); this strategy has been shown to be effective in studies of single-hydride transfer in enzymes,\cite{Boekelheide2011} electron transfer between metal complexes,\cite{Kenion2016,Menzeleev2011} and hydrogen clusters inside clathrate hydrates.~\cite{Witt2010} In addition, there have been demonstrations of mixTS even beyond the quantum-classical limit that motivate the present software development.~\cite{Ramirez1994, Herrero1995, Herrero1996, Herrero1996-2, Wallqvist1986, Kretchmer2016-2} 

    In this paper, we introduce MixPI, a driver for the CP2K software package that allows for each particle in a PIMD simulation to have a unique number of beads.
    We derive the general mixTS-PIMD Hamiltonian in Sec \ref{sec:methods} and describe its implementation and the workflow of the MixPI 
    program in Sec \ref{sec:implementation}. In Section \ref{sec:results}, we present benchmark results for two systems of interest to demonstrate the accuracy of the mixTS method and compare simulation times. We discuss the particular types of systems where the mixTS method can be expected to offer a significant advantage over existing PIMD implementations. We also briefly outline planned extensions to MixPI that will incorporate recent advances in PIMD simulation methods and also automate approximate real-time correlation function calculations using RPMD and CMD.
    
    \section{Theory} \label{sec:methods}
	We briefly review PIMD theory and define terms in the mixTS-PIMD framework following previous work.~\cite{Steele2011, Ceriotti2010} Complete derivations of the all-replica and mixTS Hamiltonians are 
 included in the MixPI User Manual.~\cite{mixpi-github}
        
		\subsubsection{All Replica Ring Polymer Hamiltonian} \label{sec:pimd_ham}
			The quantum canonical partition function can be written as
			\begin{equation}
			\begin{split}
			    Z & = \text{Tr} [e^{-\beta \hat{H}}] \\
			    & = \int dq \bra{q} e^{-\beta \hat{H}} \ket{q},
			\label{eq:partfn}
            \end{split}
			\end{equation}
			where $\ket{q}$ is a position eigenstate of the one-dimensional Hamiltonian
			\begin{equation}
			\begin{split}
				\hat{H} & = T(\hat p) + V(\hat q).\\
			\end{split}
			\end{equation}
    The trace in Eq.~\ref{eq:partfn} is evaluated by inserting $N-1$ copies of the identity in position space and using the Trotter expansion to evaluate the kinetic and potential energy matrix elements. Introducing $N$ normalized Gaussians in momentum space, we obtain
			\begin{equation}
				Z = \lim_{N \rightarrow \infty} \left( \frac{\beta}{2 \pi m_f} \right)^\frac{N}{2} \left( \frac{m N }{2\pi\beta\hbar^2} \right)^\frac{N}{2}  \int d\{\textbf{p},\textbf{q}\} 
				e^{-\beta H_\text{RP}(\{\textbf{p}, \textbf{q}\})}
                \label{eq:1d-partition-function}
			\end{equation}
   where $\beta=1/k_B T$, $T$ is temperature, and the exponent is 
   the classical, $N$-bead, ring-polymer Hamiltonian,
			\begin{equation}
				H_{\text{RP}} = \sum_{\alpha=1}^N  \left(\frac{p_{\alpha}^2}{2m_f} + \frac{1}{N}V(q_{\alpha}) + \frac{N m}{2 \beta^2 \hbar^2} (q_{\alpha} - q_{\alpha+1})^2\right),
            \label{eq:rpham}
			\end{equation}
   with neighboring beads connected via harmonic springs and $q_{N+1}\equiv q_1$. In Eq.~\ref{eq:1d-partition-function}, we use the notation $\{\textbf{p}, \textbf{q}\}$ to indicate the vector of all bead positions and momenta and similarly define $\int d\{\textbf{p}, \textbf{q}\}\equiv\prod_{\alpha=1}^N \int dp_\alpha \int dq_\alpha$. 
   For simplicity, the expressions above are written for a 1D system but can be trivially extended to higher-dimensional systems.
   Classical MD trajectories generated by the ring polymer Hamiltonian in Eq.~\ref{eq:rpham} are used to sample the canonical phase space of the quantum system in standard PIMD simulations. We note that equilibrium properties obtained in a PIMD simulation are independent of the choice of the fictitious mass, $m_f$, introduced in Eq.~\ref{eq:rpham} that is different than the physical mass, $m$.~\cite{Parrinello1984} 

		\subsubsection{MixTS Ring Polymer Hamiltonian}\label{sec:mts_ham}
           Here, we derive an expression for the mixTS ring polymer Hamiltonian of a system of three 1-D particles where each particle is mapped to a different number of beads. 
		     The quantum Hamiltonian operator for three particles is written
			\begin{equation}
			\begin{split}
                				\hat{H} & = \sum_{i=1}^3 \frac{\hat{p}_i^2}{2m_i} + \sum_{i=1}^3 V_i(\hat{q_i}) + \sum_{i=1}^3 \sum_{j>i} V_{ij}(\hat{q_i},\hat{q_j}) \\ 
                &+ \sum_{i=1}^3 \sum_{j>i} 
				\sum_{k>j} V_{ijk}(\hat{q_i},\hat{q_j}, \hat{q_k}) 
			\end{split}
			\end{equation}
			where $\hat{p_i}$, $\hat{q_i}$, and $m_i$ are the momentum, position, and mass of the $i^{th}$ quantum particle. $V_i(\hat{q_i})$ is the uncoupled single-particle component of the 
			potential energy for the $i^{th}$ degree while $V_{ij}(\hat{q_i},\hat{q_j})$ and $V_{ijk}(\hat{q_i},\hat{q_j},\hat{q_k})$ include the pair-wise particle interaction potentials and the three-body terms respectively. It is important to note that while we choose to explicitly illustrate the form of the mixTS Hamiltonian for a 3-particle system, the general equations can be extended to any number of particles and to potentials that contain more than three-body interactions.

   In deriving the mixTS Hamiltonian, we work with the most general case where the PIMD simulation numerically converges with different numbers of beads $N_i$ for different degrees of freedom $q_i$. We will assume here that $N_1\ge N_2 \ge N_3$ suggesting that the $q_1$ degree of freedom makes the largest contribution to any observed quantum effect. The three-body Hamiltonian can now be split into three component Hamiltonians, 
      \begin{align}
        \nonumber
         \hat H_1 &= T(\hat{p_1}) + V_1(\hat{q_1}) + V_{12}(\hat{q_1}, \hat{q_2})\\
         & + V_{13}(\hat{q_1}, \hat{q_3}) + V_{123}(\hat{q_1}, \hat{q_2}, \hat{q_3}),\\
         \hat H_2 &= T(\hat{p_2}) + V_2(\hat{q_2}) + V_{23}(\hat{q_2}, \hat{q_3}), \,\,\text{and}\\
         \hat H_3 &= T(\hat{p_3}) + V_3(\hat{q_3}).
         \label{eq:comp_ham}
    \end{align}
    Following our assumption that $N_3$ represents the smallest number of replicas required to converge, we use the asymmetric Trotter approximation sequentially,
	\begin{align}
            \nonumber
			Z &= \lim_{N_3, N_2, N_1 \rightarrow \infty} \int dq_1 \int dq_2 \int dq_3 \\
            &\times\left\langle q_1 q_2  q_3 \left|  \left( 
				\Lambda_3 \left( \Lambda_2 \Lambda_1^{\frac{N_1}{N_2}} \right)^{\frac{N_2}{N_3}} \right)^{N_3} 
                 \right|q_1 q_2  q_3\right\rangle,
	\end{align}
            where we define $\Lambda_i = e^{-\frac{\beta H_i}{N_i}}$, and we choose the three replica numbers such that the ratios $N_1/N_2$ and $N_2/N_3$ are integers. 
            Evaluating the high-temperature matrix elements for each component Hamiltonian we obtain an expression for the canonical partition function, 
\begin{align}
    \nonumber
   Z&\propto\lim_{N_3, N_2, N_1 \rightarrow \infty} 
   \int d\{p_1,q_1,p_2,q_2,p_3,q_3\}\\
   & \times 
   e^{-\beta H^\text{mix}_\text{RP}(\{p_1,q_1,p_2,q_2,p_3,q_3\})}
\end{align}

\begin{align}
                &H^\text{mix}_\text{RP} = H_0 + V_\text{mix} \label{eq:mts-hamiltonian-a}\\
                &H_0 = \sum_{i=1}^3\sum_{\alpha=1}^{N_i}  \frac{p_{i,\alpha}^2}{2m_i} + \frac{N_i m_i}{2 \beta^2 \hbar^2} (q_{i,\alpha} - q_{i,\alpha+1})^2 \label{eq:mts-hamiltonian-b}\\
                \begin{split}
                &V_\text{mix}= \sum_{i=1}^3 \frac{1}{N_i} \left( \sum_{\alpha=1}^{N_i} V_i(q_{i,\alpha}) \right.\\
                & + \left.\sum_{j>i} \sum_{\gamma=1}^\frac{N_i}{N_j} \left( \sum_{\alpha=1}^{N_j}  V_{ij}( q_{i, (\alpha-1)\frac{N_i}{N_j} + \gamma }, q_{j, \alpha})+ \right. \right. \\
                &\left. \left. \sum_{k>j} \sum_{\alpha=1}^{N_k} \sum_{\lambda=1}^\frac{N_j}{N_k}  V_{ijk}(q_{i,((\alpha-1)\frac{N_j}{N_k} + \lambda-1)\frac{N_i}{N_j} + \gamma}, q_{j,(\alpha-1)\frac{N_j}{N_k} + \lambda}, q_{k,\alpha}) \right) \right),
           \end{split}
                \label{eq:mts-hamiltonian}
			\end{align}
            where we use the notation $q_{i,\alpha}$ to indicate the position of bead ${\alpha}$ of the $i^\text{th}$ particle ring polymer.
            While the mixTS Hamiltonian in Eq.~\ref{eq:mts-hamiltonian} requires keeping track of indices to ensure that the correct bead numbers on different atom ring polymers interact with each other, we note that the implementation is straightforward. It also offers a significant advantage by decreasing the total number of force evaluations necessary for systems and observables where a large number of atoms can be converged with small $N$ values, and only a handful require larger bead numbers. 

        \begin{figure}[h]
            \centering
            \includegraphics[scale=0.4]{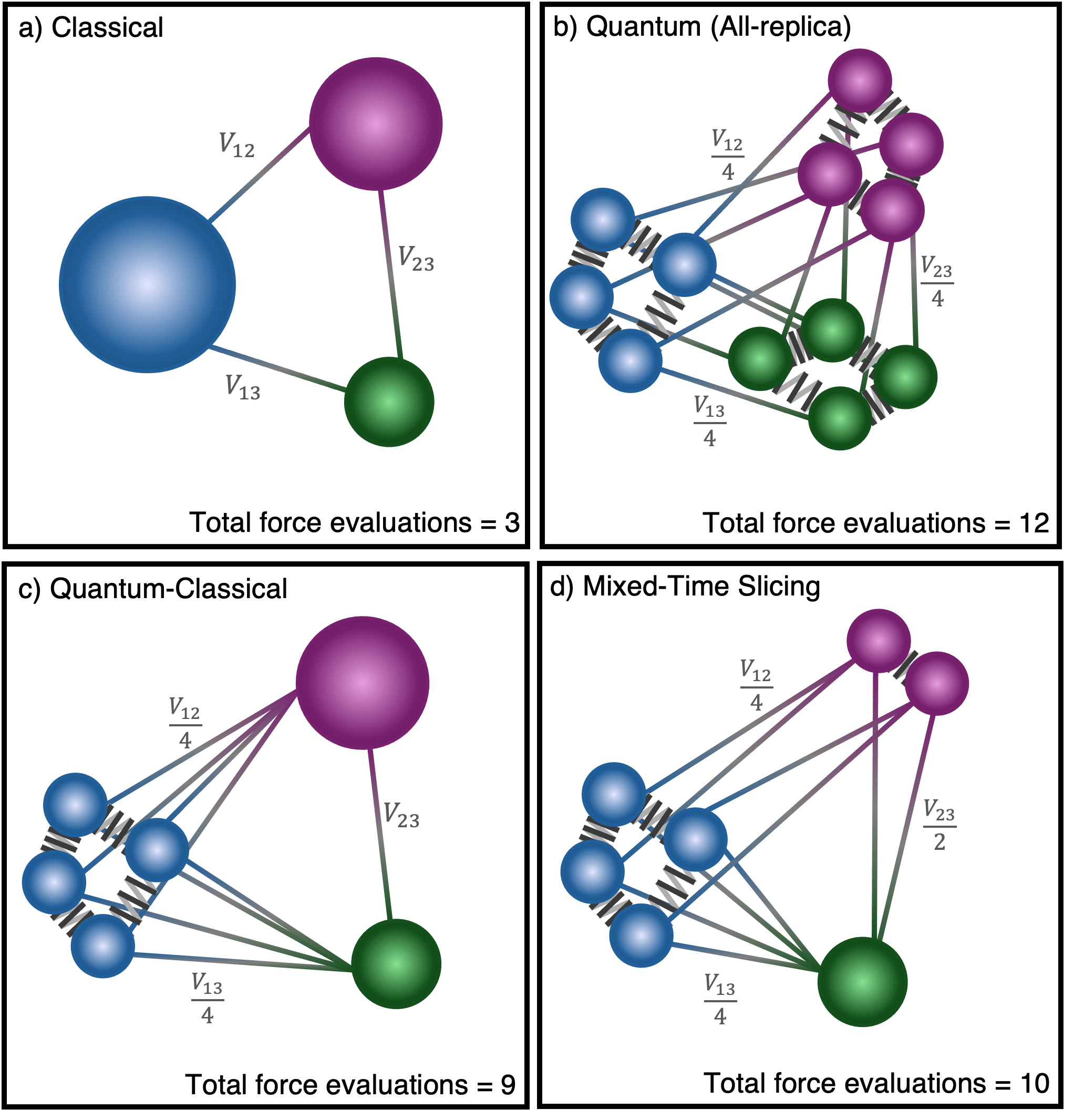}
            \caption{A diagram showing the different forces between ring polymers at four different quantization schemes: all classical (a), all quantum (b), quantum-classical (c), and mixed-time slicing (d). The size of the atoms in the cartoon reflect the extent to which we expect quantum effects to play a role in the overall system properties.}
	\label{fig:quantization-diagram}
        \end{figure}
        
        In Fig. \ref{fig:quantization-diagram}, we show the number of individual two-body forces calculated between three particles described with different bead numbers. In Fig.~\ref{fig:quantization-diagram}a, a single `classical' force is evaluated between any two particles, $V_{ij}(q_i,q_j)$. 
        If $N_1 = N_2 = N_3$, as in Fig \ref{fig:quantization-diagram}b, the potential in Eq.~\ref{eq:mts-hamiltonian} between two particles (in this case, particles 1 and 2) simplifies to $\sum_{\alpha=1}^{N_1} \frac{1}{N_1}V_{12}(q_{1,\alpha}, q_{2,\alpha})$. 
        This is the standard PIMD scheme where the inter-particle forces correspond to each bead of one particle experiencing a force from its counterpart bead for the other particle. In this example, bead 1 on particle 1 only experiences an inter-particle force from bead 1 on particle 2. This force is one-fourth the magnitude of the force between the two classical particles since $N_1 = 4$, and there are four forces evaluated between the two particles.
            
        When we have a different number of beads on different particles, the simplest case occurs when particle 2 is treated classically as shown in Fig. \ref{fig:quantization-diagram}c, so that $N_14$, $N_2=N_3=1$. 
        The two particle potential between particles 1 and 2 simplifies to $\sum_{\gamma=1}^{N_1} \frac{1}{N_1}V_{12}(q_{1,\gamma}, q_{2,1})$ where the single bead on particle 2 interacts will all the beads on particle 1. 
        In the more complicated mixTS example shown in Fig. \ref{fig:quantization-diagram}d, where $N_1=4$ and $N_2=2$, the expression for the inter-PI potential becomes $\sum_{\alpha=1}^{N_2} \sum_{\gamma=1}^{N_1/N_2} \frac{1}{N_1}V_{12}(q_{1,2(\alpha-1) + \gamma}, q_{2,\alpha})$, highlighting that beads 1 and 2 on particle 1 will interact with bead 1 on particle 2, and beads 3 and 4 on particle 1 will interact with bead 2 on particle 2. 

            We note that, in general, any multi-body force or energy calculation requires $N_{\text{max}}$ number of individual force or energy evaluations. 
            Therefore, when using potentials that contain all-body terms, like those in DFT-based methods, the all-replica results and the cost of a mixTS simulations is likely to be similar. This makes the mixTS-PIMD method most appropriate for systems that are described by additive force-fields and energy schemes. In the implementation section below, we will discuss in more detail the relative computational costs of mixTS-PIMD versus other implementations.

            In the two systems discussed below, we utilized the smooth-particle mesh Ewald routine in CP2K to analyze the long range, periodic electrostatic energies and forces.~\cite{Essmann1995} This routine utilizes a many-body expression for the electrostatic energy. Since the electrostatic forces are smoothly varying, we approximate the energies and forces between path-integral beads to be the forces between the centroids. A similar approximation has been used in previous PI techniques, such as ring polymer contraction methods,\cite{Markland2008,Marsalek2016,Kapil2016} and has been shown to return approximate forces and energies that are in close agreement with the exact PI results. The use of centroid force approximations, along with other approximate force treatments, will be explored in future work to expand the applicability of mixTS-PIMD beyond force-field treatments.

         \begin{figure*}[ht!]
            \centering
            \includegraphics[width=4.0in]{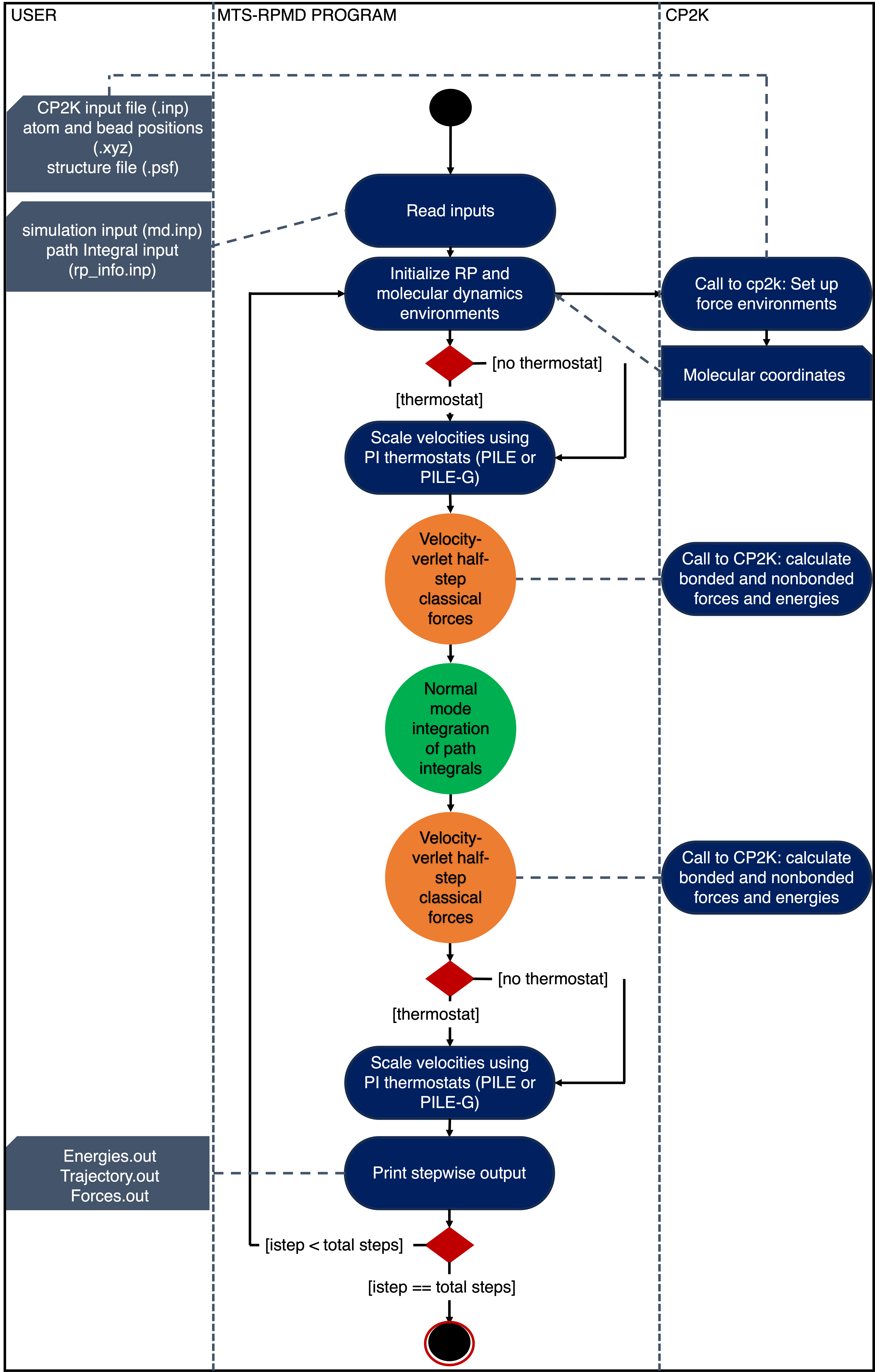}
            \caption{An activity diagram that outlines the procedure followed by the mixTS-PIMD software program MixPI. The activity diagram is partitioned into lanes to indicate the input and output files created and accessed by the user (labeled ``USER" on left), the procedure followed by the MixPI program as it proceeds through the molecular dynamics steps (labeled ``MixPI" in center) and the subsequent locations where it accesses the force and energy evaluation subroutines in CP2K (labeled ``CP2K" on right). 
            In this figure, the squares correspond to files or information that is either an input, an output, or passed from one program to another. An oval corresponds to a task, and red diamonds indicate a decision. Circles are used to show the start and end of individual subroutines.}
	\label{fig:activity}
        \end{figure*}

	\section{Implementation}\label{sec:implementation}
      This section will outline the workflow of MixPI and discuss the the required input and output for MixPI simulations. Additional instructions on compiling and running MixPI, along with a set of example calculations, are available on the MixPI software page.~\cite{mixpi-github} 

      \subsection{Overview}
		The MixPI code is implemented as a standalone program that utilizes the CP2K software
        suite \cite{cp2k} as an external library to implement mixTS-PIMD simulations. This separation allows MixPI to take advantage of the optimized
        and parallelized force and energy evaluations in CP2K while it handles the ring-polymer specific computations like normal mode evolution for the harmonic springs and book-keeping the bead indices to implement the mixTS Hamiltonian interactions. Because of this separation, limited edits were made to the CP2K source code.      

        The workflow for MixPI has some substantial changes from the workflow utilized by many of the all-replica PI software packages. In previous PI software, N copies (where N is the number of PI beads) of the system are generated and the forces for each copy of the system are independently calculated. In MixPI we generate a single system which contains all of the beads for each particle; we then generate an accurate exclusion list to account for the types of bead-bead interactions that occur in the ring polymer Hamiltonian. Fig. \ref{fig:activity} shows the activity outline of MixPI. 

        \subsection{Setup}
            To use MixPI, CP2K must be compiled as an executable and as a library. Information on how to compile CP2K for this use is available on both the CP2K installation instructions \cite{cp2k} as well as in the GitHub instructions for MixPI; the information for changes to the CP2K source code is also on the MixPI GitHub. \cite{mixpi-github}
            The necessary input files to run MixPI are separated into three categories: information on the molecular force-fields and coordinates, which are utilized by CP2K; information on the specific molecular dynamics parameters, such as temperature and time step; and information concerning ring polymer specific information like number of ring polymers, centroid constraints, and the corresponding bead numbers. 

            To specify the molecular force fields and coordinates, the standard CP2K input format for a molecular dynamics simulation is used with a few specific caveats. The coordinate file can be any accepted format; however, all molecules that contain ring polymers must be listed before any classical molecules. In the coordinate file, all of the bead positions and classical atom positions are listed. A protein structure file (PSF) is used to indicate atomic and connectivity information. Within these PSF files, the particular RP-RP and RP-classical bonding structure must be specified according to Eq. \ref{eq:mts-hamiltonian}. A specific format for the segment name, atom name, and atom type is required to indicate the presence and number of beads of each ring polymers. 
            The exact format for the PSF file is shown in the the MixPI User Manual. Within all CP2K files (force-field files, PSF, etc.), the ring polymer scaling must be integrated; for example, if using a Lennard-Jones interaction between atom A and atom B, the epsilon parameter would be scaled by $N_{\text{max}} = \text{max}(N_A, N_B)$. While it is not required to use PSF files for MixPI, one must follow the naming conventions for the atom name, atom type, and molecule name in order to accurately calculate PI forces. An input subroutine is located on the MixPI GitHub which will convert atomistic CP2K input (force-field files, PSF files, and coordinate files) into PI specific input.

            The second class of input files are related to the molecular dynamics parameters like step size and thermostat constants. This input is utilized solely by MixPI. The final class of input files are specific parameters related to the ring polymer setup and evaluation. These parameters include the number of ring polymers and the subsequent number of beads for each ring polymer. To enforce integer ratios of $N_{\alpha}$ values,  MixPI requires that all $N_{\alpha}$ values be powers of 2. The input file format is shown in detail in the MixPI User Manual. \cite{mixpi-github}
            
        \subsection{Evaluation}

           Since PIMD and related methods employ standard classical molecular dynamics trajectories, the overall flow of MixPI is similar to other MD routines. Below we highlight some of the significant deviations from traditional MD routines. 

            One of the concerns when using PIMD methods is the range of time scales needed to correctly simulate a system. In order to accurately sample configuration space in larger, condensed phase systems, one must use a time-step that allows for access to picosecond or larger time scales. However, because the mass of the ring polymer beads is small and the resulting intra-bead spring forces are large, a small time step must be used to conserve energy. A series of approaches have been developed to help overcome this limitation.
            Since the kinetic term is introduced via identity in Eq. \ref{eq:1d-partition-function}, one can select a larger fictitious mass for the beads in a simulation when evaluating static equilibrium properties. With the use of a larger fictitious mass, on the order of the lightest ``classical" mass in the system, one can overcome some of the sampling challenges which arise from the disparate timescales. 
            
            An additional modification is also made to the standard MD velocity-verlet integrator. As shown in Eq. \ref{eq:mts-hamiltonian-a}, the MixTS Hamiltonian  can be split into a free ring polymer Hamiltonian
            \begin{equation}
				H_{RP}^0  = \sum_{\alpha=1}^{N_{RP}} \sum_{\gamma=1}^{N_{\alpha}} \frac{p_{\alpha,\gamma}^2}{2m_{\alpha,f}} - \frac{N_{\alpha} m_{\alpha}}{2 \beta^2 \hbar^2} (q_{\alpha,\gamma} - q_{\alpha,\gamma+1})^2 \\
                \label{eq:free-rp-hamiltonian}
			\end{equation}
            and an external potential ($V_\text{mix}$). We can integrate the total Hamiltonian using a symplectic scheme which utilizes a split propagator
            \begin{equation}
                e^{-\Delta t L} = e^{-\Delta t/2 L_{\text{mix}}} e^{-\Delta t L_{RP}^0} e^{-\Delta t/2 L_{\text{mix}}}
            \end{equation}
            where $L_{\text{mix}}$ is the Liouvillian associated with the external potential and the $L_{RP}^0$ is the Liouvillian for the free ring polymers.~\cite{Ceriotti2010,Ceriotti2011} This splitting allows for the free ring polymer to be integrated exactly via the ring polymer normal modes and a larger time step to be utilized for normal mode propagation versus traditional velocity-verlet propagation of the free ring polymer forces.~\cite{Ceriotti2010,Ceriotti2011} Both the normal mode propagation and the manual velocity-verlet integration schemes are implemented in MixPI. Activity diagrams for the two integration schemes described above are located in the MixPI User Manual. 

        \subsection{Output and Analysis}
            After each time step, MixPI generates an updated output file that contains information about the simulation including total energy, kinetic and potential energy estimators, temperature, and the updated molecular coordinates. In its current implementation, MixPI is optimized for atomistic PIMD simulations; however, MixPI can also be used to calculate real-time RPMD correlation functions using the generated NVE trajectory files.
            The output can also be used by various PIMD/RPMD analysis codes to interpret the data and calculate observables for comparisons. Examples of this analysis include radial distribution functions (for PIMD) and diffusion constants and reaction rates (for RPMD). Future work on MixPI will involve integrated tools to perform these analyses to perform automated RPMD simulations.

\section{Examples and Discussion}\label{sec:results}
	\subsection{System I: Bulk Water}

        \begin{figure}[ht!]
            \centering
            \includegraphics[scale=0.75]{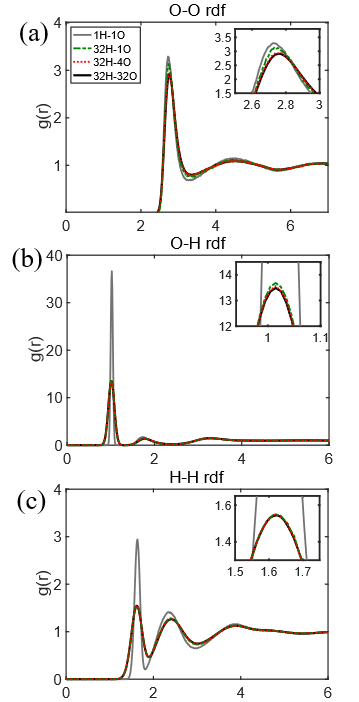}
            \caption{The O-O (a), O-H (b), and H-H (c) rdfs for a box of q-SPC/Fw water from PIMD simulations with different bead numbers. The pure classical simulation ($N_H=N_O=1$, solid grey line) predicts an over-structured water configuration compared to the fully coverged all-replica ($N_H=N_O=32$, solid black line) PIMD result. Working in the mixTS framework, we perform a quantum-classical simulation where $N_H=32$ and $N_O=1$ (dotted green line) and show that the results agree well with the all-replica PIMD simulations 
            for the H-H. However, to reproduce the O-H and O-O rdfs, we find that it is necessary to employ $N_O=4$ beads for the O atoms (dotted red line) to fully coverge to the all-replica PIMD result. }
	\label{fig:rdf}
        \end{figure}
        
        \begin{table*}
        \caption{Absolute timings for different subroutines in bulk water simulations across four quantization levels \label{table:timings}. Simulations were conducted on 64 cores across 2 nodes with hyperthreading. All simulations utilized a 0.5 fs time-step. In parantheses, we also show the scale factor in timing relative to the $N=1$ fully classical PIMD result.}
        \scriptsize
            \begin{tabular}{rcccccc}
            \hline
            \hline
            Quantization & Particles  & Thermostat & Force Evaluate  & PI-Evaluation  & One-step & Total Simulation (1 ps)  \\
            \hline
            Classical $N_H = 1$, $N_O = 1$   & 270   & .00800    & .00400 & -- & 0.0310  & 63.840\\
            Quantum-Classical $N_H = 32$, $N_O = 1$ & 5850 (22)  & 0.179 (22)    & 0.032 (8)    & 0.012 (1) & 0.504 (16)    & 1194.662 (19) \\
            Mixed-Time Slicing $N_H = 32$, $N_O = 4$   & 6120 (23)  & 0.180 (22) & 0.044 (11)  & 1 (1) & 0.563 (18)   & 1331.835 (21) \\
            Quantum $N_H = 32$, $N_O = 32$  & 8640 (32)  & .380 (47) & 0.266 (72)  & 0.0770 (6)   & 0.857 (27)   & 2275.957 (35) \\
            \hline
            \hline
            \end{tabular}
        \end{table*}

        \begin{figure}[ht!]
            \centering
            \includegraphics[scale=0.73]{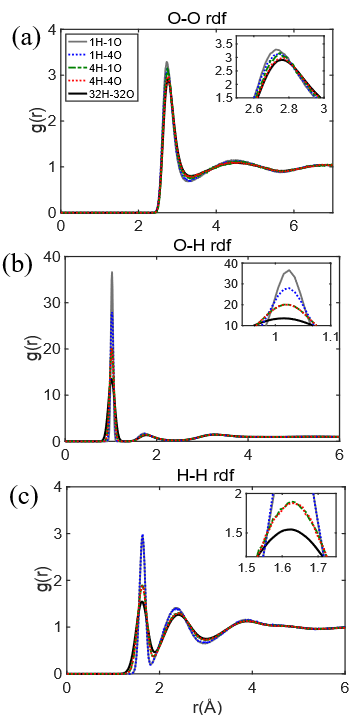}
            \caption{The O-O (a), O-H (b), and H-H (c) rdfs for a box of q-SPC/Fw water for three different mixTS treatments focused on capturing the exact O-O rdfs. The classical results ($N_H=N_O=1$, solid gray) and all-replica results ($N_H=N_O=32$, solid black) are shown for reference. We start with a counter-intuitive case using $N_H=1$ and $N_O=4$ (dotted blue) and find that while the results are reasonably accurate for the O-O rdf but quite inaccurate for the $H-H$ and $O-H$ rdfs. We also show $N_H=4$ and $N_0=1$ results (dotted green) that proves more accurate for the O-H and H-H rdfs. Finally, we show a low-bead all-replica $N_H=N_O=4$ bead simulation (dotted red) that proves exact for the O-O rdf but remains inaccurate for the other two rdfs.}
	\label{fig:rdf-OH-partial-quantize}
        \end{figure}

       Bulk water has been studied extensively with PIMD and RPMD methods due to the importance of NQEs on the structure and dynamics of water. \cite{Miller2005,Kuharski1985, Wallqvist1985,Habershon2009,Habershon2011,Ceriotti2013, Ceriotti2016}
       Because of the wealth of previous results, the radial distribution functions (rdfs) for bulk water are used here to test the functionality of the MixPI program and to establish systematic convergence in the PIMD simulations allowing different numbers of beads for the H and O atoms. 
       All simulations here are performed with a 14 $\dot{A}$ box of 90 water molecules with periodic boundary conditions using the q-SPC/Fw water model which is parameterized for use with ring polymers to prevent over-inclusion of NQEs. ~\cite{Paesani2006,Habershon2009,Habershon2011,Machida2017} 
       The rdfs for each system are computed from the average of 5 independent 20 ps NVT simulations for the mixTS and all-replica PIMD simulations with a path integral Langevin thermostat (PILE).~\cite{Ceriotti2010} The classical rdf is obtained from a single trajectory of length 500 ps. 

        In Fig. \ref{fig:rdf}, we show the rdfs for bulk water using four different quantization schemes. For the classical and mixTS simulations, configurations were recorded every 100 fs. The classical result is obtained using $N_H=N_O=1$, where $N_H$ and $N_O$ are the number of beads used for the H and O atoms respectively. This result is the most different from the fully converged, all-replica, quantum result obtained with $N_H=N_O=32$ beads, in agreement with previous work.~\cite{Habershon2009,Paesani2006} Comparing the quantum-classical result with $N_H=32$ and $N_O=1$ with the exact quantum result, we find that while the H-H rdfs agree well, there are small deviations in the O-H and O-O rdfs. The difference in number of beads required to achieve convergence provide an interesting insight: while quantizing the O atoms does not affect the H-H rdf, the O-H and O-O rdfs involve NQE contributions from the O atoms as well. We confirm this hypothesis by performing a mixTS simulation with $N_H=32$ and $N_O=4$ and show that it is able to reproduce the all-replica PIMD result exactly for the O-H and O-O rdfs as well as the H-H rdf. 
        
        The absolute and relative timings for our simulations are listed in Table \ref{table:timings}; these simulations were conducted on two 32-core nodes with hyperthreading enabled (AMD EPYC 7502). The relative timings are all referenced with respect to the classical $N_H=N_O=1$ results. As expected, the larger the number of beads, the longer the simulation time. The timing breakdown demonstrates clearly that the most significant increase comes from the need to evaluate the inter-particle inter-bead forces. While there is also a small increase in time for the thermostat and the normal model ring polymer propagation at higher $N$ values, these increases are typically negligible compared to the overall cost of force evaluations. 
        
        We further investigate the idea that different observables (rdfs in this case) converge with different levels of quantization. Specifically, we investigate if the O-O rdfs converge with fewer bead numbers, particularly for the H atoms. Figure~\ref{fig:rdf-OH-partial-quantize} demonstrates this is indeed the case; we can fully converge the O-O rdf using $N_H=N_O=4$ and get reasonably close agreement with the $N_O=4$, $N_H=1$ result, but we see significant inaccuracies in both O-H and H-H rdfs because of the failure to include NQEs due to the H-atoms. These results together demonstrate that mixTS is not only a path to reducing the computational cost of PIMD simulations but that it can also be used to establish the importance of types of NQEs for the calculation of specific observables.

	\subsection{System II: Aqueous Co$^{2+}$}

        \begin{figure}[h!]
            \centering
            \includegraphics[scale=0.8]{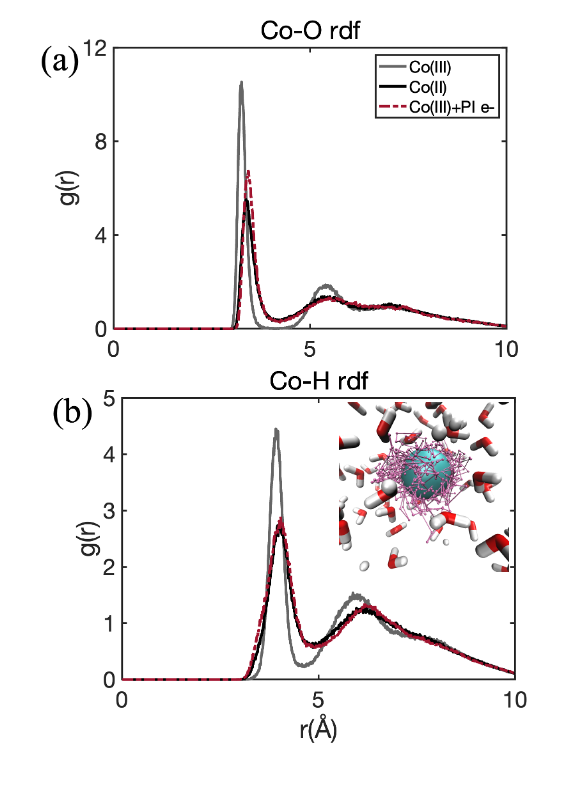}
            \caption{The Co-O (a) and Co-H (b) rdfs for aqueous Co-ion systems. We show two classical results: aqueous Co$^{2+}$ (black) and Co$^{3+}$ ions (gray). We compare these rdfs to the result for a Co$^{3+}$ ion with an electron treated as a 1024 bead ring polymer centered on the ion (dotted green). We see relatively good agreement between the Co$^{2+}$ and Co$^{3+} + e^-$ rdfs. In (b), a snapshot from the Co$^{3+} + e^-$ trajectory is shown with the Co$^{3+}$ ion in blue and the beads for the e$^-$ shown in pink. 
            }
	\label{fig:rdf_co}
        \end{figure} 
 
        Prior quantum-classical PIMD/RPMD simulations have studied the structure and dynamics of electron transfer in metal centers and organometallic complexes. 
        \cite{Kenion2016,Menzeleev2011,Kretchmer2016-2}
        The original simulations were performed using in-house code developed by the researchers; here we demonstrate that 
        MixPI can implement similarly high-dimensional quantum-classical
        simulations that will find application in the study of 
        charge transfer reactions.
        We demonstrate MixPI's functionality using solvated Co$^{2+}$ ion as a small, test system.~\cite{Kenion2016} 
        We compare the rdfs for the system across treatments: classical Co$^{2+}$ in a box of classical water, classical Co$^{3+}$ in a box of classical water, and classical Co$^{3+}$ with a RP electron centered on the metal cation (a $1024$-bead electron ring polymer) in a box of classical water. The simulations are performed with the cation centered in a 14 $\dot{A}$ box of 91 water molecules to match the density of bulk water at room temperature. 
        Each simulation was conducted using a Lennard Jones potential ($\epsilon = 666.77 \text{ } K_e$ and $\sigma = 4.134 \text{ }\dot{A}$) for the Co - water interactions and the SPC/Fw water force field. 
       For the fully classical simulations, four independent trajectories of 50 ps each with a 0.5 fs time step where conducted with configurations recorded every 25 fs.
       For the Co$^{3+}$ + e$^{-}$ simulation, we used four independent trajectories of 15 ps each with a 0.01 fs time with configurations recorded every 5 fs. 
       The rdfs from the independent trajectories were averaged and are shown in Fig.~\ref{fig:rdf_co}; the resulting error bars are within the line widths shown.
        As expected, the Co$^{2+}$ and Co$^{3+} + e^{-}$ Co-O and Co-H rdfs are similar; the Co$^{3+}$ + e$^{-}$ simulation is able to capture the water rearrangement in the first solvation shell to mimic the water structure of Co$^{2+}$ (when compared to the Co$^{3+}$ solvation structure). The small differences in the Co-O rdf between Co$^{2+}$ and Co$^{3+} + e^{-}$ may be due to the delocalized nature of the PI electron as shown in the insert in Fig. \ref{fig:rdf_co}b. 

        In Table \ref{table:co_timings}, we report timings for individual subroutines in the Co$^{3+} + e^-$ simulation with relative timings compared to the $N=1$ classical simulation. When comparing the timings to those in Table \ref{table:timings}, we note that while the Co$^{3+} + e^{-}$ simulation has a faster force evaluation than the all-replica quantum water simulation, the time it takes to propagate the electron RP becomes increasingly significant due to the large $N$ value. It is also worth noting the total simulation time is much larger for the Co$^{3+}+e^{-}$ system due to the smaller time step needed to conserve energy.

        \begin{table}
        \centering
            \begin{tabular}{|c|c|}
            \hline
            Subroutine & Timings \\
            \hline
            Particles & 1298 \\
            Thermostat & 0.085 (11) \\
            Force Evaluate  & 0.106 (35) \\
            PI-Evaluation  & 0.0540 \\
            One-step  & 1.0660 (36)\\
            \hline
            \end{tabular}
        \caption{Absolute timings for subroutines in the Co$^{3+}$ + 1024-bead e$^-$ RP simulation conducted on 64 cores across 2 nodes with hyperthreading. A 0.01 fs time step is employed to ensure energy conservation. In parentheses, we show the relative timings compared against the simulation of classical Co$^{2+}$ in water.}
            \label{table:co_timings}
        \end{table}

\section{Conclusion}\label{sec:conclusion}
We introduce an atomistic code for mixTS path integral simulations, MixPI, that allows for individual atoms in a system to be described by different numbers of beads within the PI framework. We demonstrate applications of this software across two example systems, bulk water and aqueous Co$^{2+}$. We show that MixPI reproduces radial distribution functions for these systems in keeping with previous results and allows us to carefully investigate numerical convergence properties that are observable-specific. We further show that the mixTS method reduces computational time significantly for systems with a large number of atoms where only a few contribute significant NQEs. The current version of MixPI is available on GitHub \cite{mixpi-github} and interfaces with CP2K. Moving forward, we will incorporate ring polymer contraction and MTS schemes to MixPI, and develop a suite of subroutines to calculate common properties of interest including correlation functions, radial distribution functions, spectra, and reaction rates. In terms of methodology, we also plan to expand MixPI to automate real-time RPMD simulations, include additional functionalities such as an isobaric-isothermal NPT algorithm, thermostatted tRPMD, and extend it to the study of multi-level systems.

    \section{Acknowledgments} \label{sec:acknowledgements}
         The authors are grateful to Greg Schenter for useful discussions. B.A.J. and C.J.M acknowledge support by the DOE Office of Science, Office of Basic Energy Sciences, Division of Chemical Sciences, Geosciences, and Biosciences, Condensed Phase and Interfacial Molecular Science program, FWP 16249. N.A. and B.A.J. acknowledge support from the U.S. Department of Energy, Office of Basic Energy Sciences, Division of Chemical Sciences, Geosciences and Biosciences under Award DE-FG02-12ER16362 (Nanoporous Materials Genome: Methods and Software to Optimize Gas Storage, Separations, and Catalysis). S.B. acknowledges support from Cornell University, Department of Chemistry and Chemical Biology and the New Frontiers Grant from the College of Arts and Science.


%
%

%


\bibliography{bibtex-pi}


\end{document}